\documentstyle[epsfig,preprint,tighten,aps]{revtex}

\begin{document}
\begin{center}

\Large

{\bf Rupture of multiple parallel molecular bonds
under dynamic loading}

\large \vspace{0.5cm}
{\sl  Udo Seifert} \vspace{0.5cm}
\normalsize

 Max-Planck-Institut f\"ur Kolloid- und
Grenzfl\"achenforschung,\\ Am M\"uhlenberg 2, 14476 Golm, Germany
 \vspace{0.5cm}
 \end{center}
\begin {abstract}

Biological adhesion often involves several pairs of specific
receptor-ligand molecules. Using rate equations,
we  study theoretically the rupture of such
multiple parallel bonds under dynamic loading assisted by thermal activation. 
For a simple generic type of
cooperativity, both the rupture time and force
exhibit several different scaling regimes.
The dependence of the rupture force on the number of bonds 
  is predicted to be 
either linear, like a square root or  logarithmic.

\end{abstract} 
\vspace{0.5cm}
\vspace{0.5cm}
\def\beq{\begin{equation}}
\def\ee{\end{equation}}
\def\pcite{\protect\cite}
\def\pa{\partial}
\def\m{{\bf m}}
\def\q{{\bf q}}
\def\l{{\cal L}^\dagger}
\def\a{{\cal A}}

\def\lsim {\protect
\raisebox{-0.75ex}[-1.5ex]{$\;\stackrel{<}{\sim}\;$}}

\def\gsim {\protect
\raisebox{-0.75ex}[-1.5ex]{$\;\stackrel{>}{\sim}\;$}}

\def\lsimeq {\protect
\raisebox{-0.75ex}[-1.5ex]{$\;\stackrel{<}{\simeq}\;$}}

\def\gsimeq {\protect
\raisebox{-0.75ex}[-1.5ex]{$\;\stackrel{>}{\simeq}\;$}}

\def\u{{\bf u}}
\def\v{{\bf v}}
\def\r{{\bf r}}
\def\ex{{\bf e}_x}
\def\ez{{\bf e}_z}
\def\p{{\partial}}
\def\n{{\bf \nabla}}
\def\gd{{\dot \gamma}}
{\it Introduction.} 
Single molecule force  spectroscopy  has made it  possible to measure 
the  binding strength of a   pair of receptor-ligand 
(``lock-key'') molecules using
vesicles \cite{z:evan91},  the atomic force apparatus 
\cite{z:flor94,z:moy94,z:lee94}, or optical
tweezers \cite{nish95} as
transducers. Thus, the essential constituents mediating biological
adhesion have become accessible to quantitative physical experiments
\cite{bong99}.
 This experimental progress has fostered theoretical studies of
 the rupture of such pairs under dynamic loading.  Thermal activation 
being a main
contributing factor, 
Kramers-like descriptions of the rupture process
 with time-dependent potentials show that the rupture strength
of such bonds depends on the loading rate \cite{z:evan97d,z:izra97,shil98}. 
Such behavior has been found experimentally indeed \cite{merk99,sims99}.
While unspecific theoretical  models of
the rupture process reveal generic features, molecular dynamic 
studies can  address the
details of the dynamics of the
rupture of specific pairs \cite{z:izra97,z:grub96}. 

Adhesive contact and the rupture thereof often involves not just one
but several  molecular pairs of the same or different species
\cite{bell78}. 
The equilibrium properties of the cooperative effects of such 
specific interactions are well studied both in theory
\cite{bell84,zuck95,lipo96} and in experiments \cite{nopp96,albe97,ches98}. 
Concerning the dynamics of  rupture of such a contact under loading, 
 detailed  models for specific  problems such as the
peeling of a membrane \cite{demb88,ra99} 
or the rolling of leucocytes in shear flow \cite{hamm87}
have been  solved numerically to extract a critical tension or shear rate for
rupture. However,  it
is inherently difficult to separate generic dynamical 
properties from specific ones
using such intricate models. As an example for  a generic property
consider the following question: How does the time and force necessary
to break  an adhesive contact under dynamic loading
 depend on the number of bonds initially present?

The present study addresses this question within a simple model that
extends work on the dynamic failure  of a single bond to that of a whole
patch involving several bonds of the same type. Quite generally,
two different limiting  cases must
  be distinguished.  If the load is primarily concentrated on
one bond at a time with relaxation of the load when the first bonds 
fails and subsequent loading of the next  one, 
 the rupture process  basically is a sequence of
similar single molecule events. The $N_0$ bonds initially present then
act {\it in series}. The rupture time will be $\sim N_0$ whereas the
force will exhibit  a saw tooth-like pattern with a peak given by
the rupture strength of a single bond. Such a behavior has been found 
and modeled in the related case of unfolding of proteins with several
identical domains like titin \cite{rief98}.
The main purpose of the present paper is
to analyze the other case where the load is distributed (almost) uniformly
among several bonds such that these bonds act {\it in  parallel}.
As more and more  bonds rupture, the force 
on the remaining ones increases. This simple type of 
cooperativity leads to different scaling regimes for the rupture time and
rupture force.

\epsfig{file=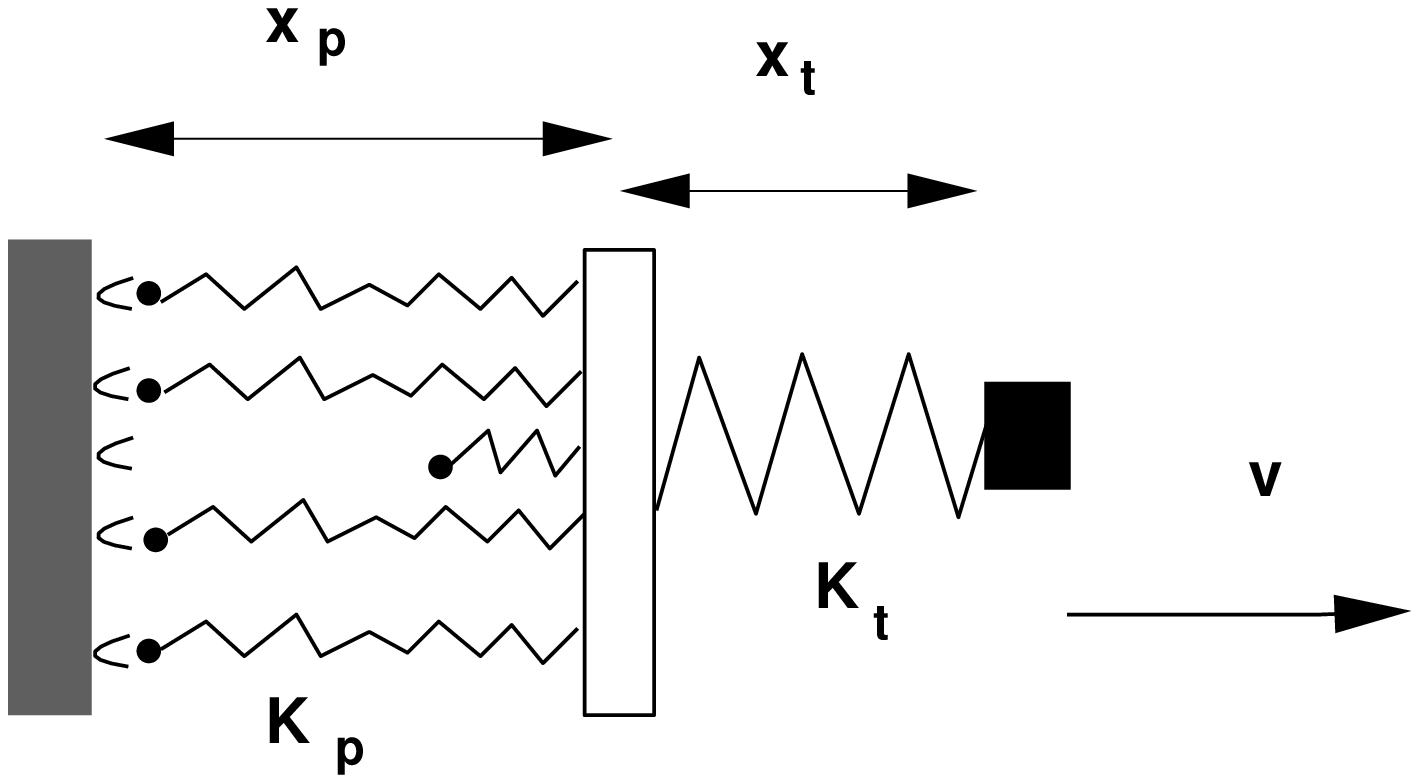, height = 5cm}

{FIG.1: Model geometry for the rupture of parallel bonds. Symbols
are explained in the main text.} 
\vskip 1cm

\def\mst{\tilde \mu_s}
\def\mht{\tilde \mu_h}
\def\ms{ \mu}
\def\mh{\bar \mu}
{\it Model.} We model the rupture geometry generically as shown in Fig.1.
One partner of the bond (``receptor'') is  confined to a substrate.
The other (``ligand'') is connected by a
polymer to a transducer which is connected by an elastic element to a
sled being pulled at velocity $v$.  
For simplicity, we model both the elasticity of the transducer
and the polymers as Hookean springs with zero rest length
and spring constants $K_t$ and $K_p$, respectively.
 As long as a bond is intact the corresponding
polymer is
stretched to an extension $x_p$ which we assume to be the same for 
all intact bonds. The elongation of the transducer from
its resting position is $x_t$. Force balance on the transducer becomes
$NK_px_p=K_tx_t$
where $N$ is the number of intact bonds.  Geometry dictates the time dependence
$ x_p(t)+x_t(t) = vt.$
From these two relations, we find the time-dependent 
force on an intact bond as 
\beq
F_b(t)=K_px_p(t)={K_pK_t\over N(t)K_p+K_t} v t.
\label{eq:F}
\ee
Following Bell \cite{bell78},
we assume that the main effect of such a  force is to introduce an
instantaneous, time-dependent dissociation rate $k_0(t)$ according to
\beq
k_0(t)=k_0\exp[F_b(t)x_b/k_BT] ,
\ee 
where $k_0$ is the dissociation rate in the absence of a force.
The quantity  $x_b$ is of the order of the distance between the minimum of
the binding potential and the barrier and $k_BT$ is the product of
Boltzmann's constant and temperature.

We are mainly interested in the case of a {\it soft}  transducer defined as
$K_t\lsim K_p$. In this case,  eq. (\ref{eq:F}) shows that the 
force on a bond is
inversely proportional to the number of intact bonds for all
$N(t)$. Hence, when a bond
ruptures, the force on the remaining ones increases.
We now discuss two different cases, irreversible and reversible bonds.
In the former case, a bond, once ruptured, cannot rebind. Reversible
bonds
have a non-zero rebinding rate.

{\it Irreversible bonds.}
Initially  $N(t=0)\equiv N_0$ bonds are present. 
The rate equation for
their time-dependent decrease is
\beq
\partial_tN=-N(t)k_0\exp [F_b(t)x_b/k_BT]  . 
\ee
We scale time with the dissociation rate in equilibrium $k_0$ 
according to 
$
\tau\equiv tk_0.$ The rate equation in the case of a
 soft transducer then becomes
\beq
\partial_\tau N=-N\exp[(\ms\tau/N]
\label{eq:rate-soft}\ee
with the loading parameter
\beq
\ms\equiv K_tx_bv/k_BTk_0  
\label{eq:ms}.\ee
This simple rate equation  seems not to have an analytical solution. 
However, its scaling behavior can be extracted by the following analysis.
With the substitution 
$
u(\tau)\equiv \tau/N $
one obtains\beq
\partial_\tau u = u(1/\tau  +1) + u(\exp[\ms u]-1)
\label{eq:rate-soft2}\ee
For small $\tau$, $u(\tau)\approx \tau/N_0$ and the second term
in (\ref{eq:rate-soft2})
can therefore be neglected.
 The solution $u_1(\tau)$ of the
corresponding equation becomes $u_1(\tau)=\tau e^\tau/N_0$ and
hence a purely exponential decay for the number of intact bonds,  
$N(\tau) = N_0e^{-\tau}$. This approximation breaks down for $\tau\gsim\tau_1$
with $\tau_1$ implicitly defined by 
\beq
{\rm max}(1/\tau_1,1) = \exp[\ms u_1(\tau_1)]-1  .\ee
 For  $\tau>\tau_1$, we can then ignore both the first term and the 
``-1'' in the second term of (\ref{eq:rate-soft2}). 
The corresponding equation 
$
\partial_\tau u = u \exp[\ms u]
$
is solved
by 
\beq
E(\ms u_1)-E(\ms u)=\tau-\tau_1
\label{eq:E}
\ee
where $E(x)\equiv \int_x^\infty dx'e^{-x'}/x'$ is the 
exponential integral and $u_1\equiv u_1(\tau_1)$ is the cross over
value of the first solution at the matching point $\tau_1$. 
Hence the time necessary for complete
rupture, $\tau^*$, can be estimated by setting $u(\tau^*) = 
\tau^*/N = \infty$ which
leads to
\beq
\tau^* =\tau_1+\tau_2= \tau_1 + E(\ms u_1).
\ee
Based on this approximative solution of (\ref{eq:rate-soft2}),
three sub-regimes can be identified: 

(i) $\ms \lsim 1$: In this case, the exponential decay holds  till
$N(\tau)\simeq 1$. Physically, the rupture is then complete. In this
trivial regime, where the loading is too small to affect the rupture process 
at all,
the time required for rupture is
\beq
\tau^* \sim \ln N_0 .
\ee Note that the same result could have been obtained
by analyzing the mean time 
required for the irreversible decay of $N_0$ independent bonds under no force.

(ii) $1\lsim \ms \lsim N_0$: In this regime, the exponential decay persists till
$\tau_1 \sim \ln(N_0/\ms)$. At this time the number of bonds has
reached $N(\tau_1)\sim \ms$. The remaining bonds decay according to
(\ref{eq:E})  which leads to an additional time $\tau_2$ 
of order 1 which is
 small compared to
$\tau_1$. Hence the whole rupture time in this regime is of
order
\beq
\tau^* \sim \ln(N_0/\ms)  .
\ee

(iii) $N_0 \lsim \ms$. In this case, the exponential decay applies till
$\tau_1 \sim (N_0/\ms)\ln(\ms/N_0)$. According to
(\ref{eq:E}) the remaining time $\tau_2 \sim 
(N_0/\ms)/\ln(\ms/N_0)$  is   smaller
than $\tau_1$. Hence the total rupture time is 
\beq
\tau^* \sim (N_0/\ms)\ln(\ms/N_0) \label{eq:tau3} .
\ee

Thus we find for small loading rates that the rupture time is
logarithmic in the number of bonds initially present whereas for
large loading, this time becomes linear in $N_0$. For fixed $N_0$
and increasing $\ms$,
the rupture time first is independent of $\ms$. It then 
decays  logarithmically  in $\ms$ and finally becomes
inversely proportional to $\ms$.

\def\ft{\tilde F}
The  force  measured by the transducer
is given by
\beq
F_t\equiv N(t) K_p x_p(t)\approx (k_BT/x_b)\ms \tau. \ee 
Thus, the total force experienced by the soft transducer is independent
of the number of intact  bonds and increases linearly in time. 
The dimensionless rupture force $f^*=\mu \tau^*$ is thus given by 
\begin{eqnarray}
f^* &\sim & \ms \ln N_0 \ \ \ \ \ \ \ {\rm for} \ \ \ms\lsim 1 , \cr
&\sim& \ms \ln(N_0/\ms) \ \ \ {\rm for} \ \ 1 \lsim \ms \lsim N_0  , \cr
&\sim & N_0\ln (\ms/N_0)  \ \ {\rm for} \ \ N_0\lsim \ms. 
\end{eqnarray}
in the three regimes, respectively.

{\it Reversible bonds.}
\def\Ne{N_{eq}}
So far, we have neglected the possibility that broken bonds can reform. Hence,
rupture from a genuine equilibrium situation where bonds form, break, and 
rebind
requires a refined  description where we add a term
for rebinding. We assume that one species
of the receptor/ligand couple is
limited to a total number $N_1$ with $N(t)$ molecules bound and
 $N_1-N(t)$ unbound whereas the other species is available in excess. The rate
equation becomes
\beq
\partial_tN=-N(t)k_0\exp [\ms\tau/N(t)] + k_f(N_1-N(t)) ,
\ee
where we assume for simplicity that the rate $k_f$ for bond formation is not
 affected by
the force. Without loading, the equilibrium number of bonds is
\beq
\Ne = \gamma N_1 /(1+\gamma)
\ee
where $\gamma\equiv k_f/k_0$. As loading starts, the number of bonds decreases
from this equilibrium value. 
With 
$
u(\tau)\equiv \tau /N
$ as before, 
we get \beq
\partial_\tau u = u(1/\tau +1 +\gamma -\gamma N_1u/\tau) + u 
(\exp[\ms u] -1) 
\label{eq17}\ee
For $\ms=0$, this equation is solved by 
\beq
u_0(\tau)\equiv \tau/\Ne,
\ee
which corresponds to the stationary equilibrium distribution.
The loading term becomes relevant at a time $\tau =\tau_1$ for which 
\beq
(\exp[\ms u_0(\tau_1) /N_1] -1) u_0 \sim  \partial_\tau u_0 =1/\Ne.
\ee
Two cases must then be distinguished:

(i) For $\ms \lsim \Ne$, $\tau_1\sim (\Ne/\ms)^{1/2}$. Up to this time, the
loading has not significantly affected the number of bonds. The remaining time
till all bonds are ruptured can be estimated to be of the same order 
as $\tau_1$ using (\ref{eq17}). Hence,
\beq
\tau^*\sim  (\Ne/\ms)^{1/2}.
\label{eq:tau-rev} 
\ee In this case, the rupture time increases as
a square root of the equilibrium bonds present and decreases as a square root
of the loading parameter.

(ii) For $\Ne \lsim \ms$, $\tau_1\sim (\Ne/\ms)\ln(\ms/\Ne)$, with a remaining time
of the same order. Hence in this case, we recover the irreversible result
(\ref{eq:tau3}) with $N_0$ replaced by $\Ne$.

Since in both cases the rupture time $\tau^*\sim \tau_1$, we get easily for the
rupture force 
\begin{eqnarray}
f^*=\ms\tau^*&\sim&(\ms\Ne)^{1/2} \ \ \ {\rm for} \ \  \ms \lsim \Ne, 
\label{eq:f-soft} \cr
&\sim & \Ne\ln(\ms/\Ne) \ \ \ {\rm for} \ \  \Ne  \lsim \ms . 
\end{eqnarray}

{\it Stiff transducer.}
So far, we have considered the case of a soft transducer $(K_t\lsim K_p)$
for which the force on a bond depends on the number of bonds.
Another limiting case is a stiff transducer with $K_t \gsim N_{0} K_p$
and $K_t \gsim N_{eq} K_p$ for the case of irreversible and
reversible rupture,  respectively.
According to eq. (\ref{eq:F}), the force on a bond  then is (almost)
independent
of the number of bonds. Hence, the rupture time is only weakly dependent
on the number of bonds. An analysis of the corresponding rate
equations along similar lines as above shows for the {\it irreversible}
case two subregimes with
\begin{eqnarray}
\tau^* &\sim&\ln  N_0 \ \ \ {\rm for} \ \ \mh \lsim  1 ,\cr
&\sim& \ln\mh/\mh \ \ \ {\rm for} \ \ \mh \gsim  1
\end{eqnarray}
with a loading parameter  
\beq
\mh\equiv K_px_bv/k_BTk_0
\label{eq:mh}\ee
 dominated by the 
polymeric stiffness. 
For the dimensionless maximal force experienced by the transducer during
the rupture process one finds 
\begin{eqnarray}
f^* &\sim&\mh N_0  \ \ \ {\rm for} \ \ \mh \lsim 1\ ,\cr
&\sim& N_0\ln \mh \ \ \ {\rm for} \ \ \mh \gsim 1 
\end{eqnarray}
in the two cases.

Similarly, for a stiff transducer and {\sl reversible} bonds, one gets
\begin{eqnarray}
\tau^* &\sim&\ (\ln  N_{eq})^{1/2}/\mh^{1/2} \ \ \ {\rm for} \ \ \mh \lsim  1 ,\cr
&\sim& \ \ln\mh/\mh \ \ \ \ \ \ \ \ \ {\rm for} \ \ \mh \gsim  1
\end{eqnarray}
and for the dimensionless maximal force experienced by the transducer
\begin{eqnarray}
f^* &\sim&\mh^{1/2} N_{eq}  \ \ \ {\rm for} \ \ \mh \lsim 1\ ,\cr
&\sim& N_{eq}\ln \mh \ \ \ {\rm for} \ \ \ \mh \gsim 1 .
\end{eqnarray}

Finally,  there is a crossover regime for $N_{0,eq}K_p \gsim K_t \gsim
K_p$, where the
pulling starts as in the soft case. As the number of intact bonds
decreases towards the value $\tilde N\equiv K_t/K_p$, the denominator
in (\ref{eq:F}) becomes  dominated by $K_t$ and the rupture process
proceeds as for
a stiff transducer. For the {\it reversible} case,
it turns out that both the rupture time and the rupture force are
dominated by the soft part. Hence the results
(\ref{eq:tau-rev},\ref{eq:f-soft})
apply for all 
$K_t  \lsim \Ne K_p$. For the {\it irreversible} case, analysis of the
crossover regime is slightly more involved. The different scaling regimes
for rupture time and force are  shown in Fig. 2 without
explicit derivation.

{\it Concluding perspective.}
Based on an analysis of rate equations,
the comprehensive
scaling analysis presented in this paper has revealed several different
regimes for the rupture time and force of parallel molecular bonds
under dynamic loading. The most distinctive regime is presumably the
square root dependence of rupture time and force  
(\ref{eq:tau-rev},\ref{eq:f-soft})  on loading 
rate and number
 of bonds derived for
reversible bonds under small loading.
Such a square root behavior on the loading rate is different from 
both the irreversible  case and the dependence on loading rate
for rupture of a single
bond or bonds in series. An experimental result showing such an exponent could 
therefore
be taken as a signature of breaking multiple parallel
reversible bonds. Of course,  it will  be important to work with
 a model system
where the number or density of bonds of at least one partner
 can be controlled in order to extract the 
dependence of rupture time and force on this crucial quantity.

An obvious theoretical refinement of the present model would be to include 
fluctuations of the rupture time for individual bonds. 
Other  ramifications can include  allowing  
lateral interactions between
the bonds,  combining 
 the simplistic Hookean transducer with a membrane patch
 with its own elasticity, or   modeling the rupture process more delicately than
done here to name just a few possibilities. 
 It will be interesting to see
how robust the scaling regimes derived in this paper will be  under such
modifications which can effectively lead to scenarios somewhere  between the
present  ``in parallel'' case and the ``in series''
case described briefly in the introduction.
 Finally, it should be clear that
in spite of   -- or rather because of -- the progress made in understanding
the single bond
behavior, the cooperative effects of several bonds under dynamic loading
deserve further attention both in theory and in experiment.

{\it Acknowledgments:} I thank E. Sackmann for a stimulating discussion and
 J. Shillcock for a critical reading of the 
manuscript.

\epsfig{file=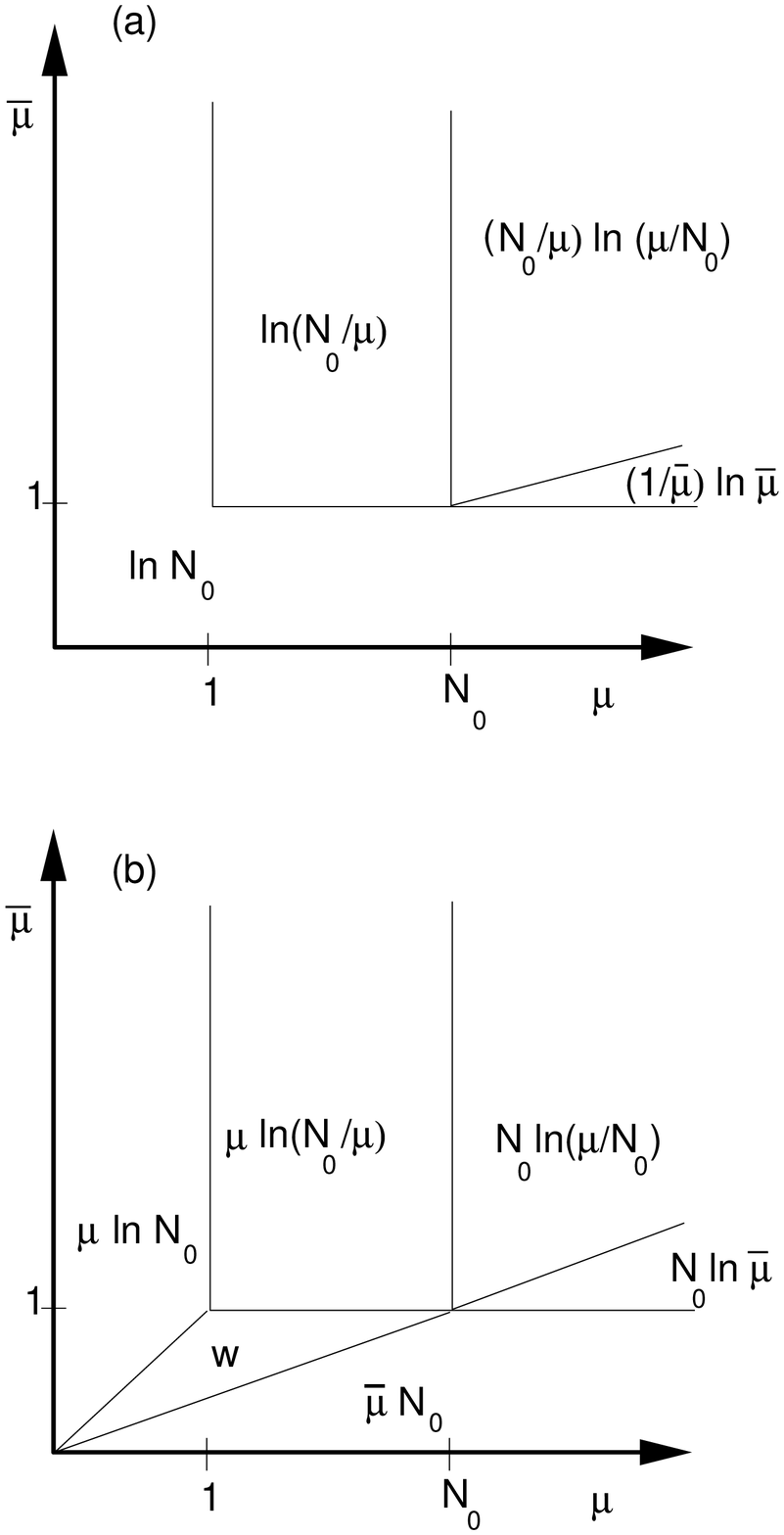, height = 17cm}

{FIG.2: Dynamical phase diagram for (a) the dimensionless
rupture time $\tau^*$ and (b) the dimensionless rupture force
$f^*$ as a function of the two loading parameters $\mu$ (\ref{eq:ms})
 and $\bar \mu$ (\ref{eq:mh}) 
in the case of  irreversible rupture. In the region $w$, the rupture force is given by
$f^*\sim \mu \ln(N_0\bar \mu/\mu)$.} 


\vskip 1cm


\begin{thebibliography}{10}

\bibitem{z:evan91}
E. {Evans}, D. {Berk}, and A. {Leung}, Biophys. J. {\bf 59},  838  (1991).

\bibitem{z:flor94}
E.-L. {Florin}, V.~T. {Moy}, and H.~E. {Gaub}, Science {\bf 264},  415  (1994).

\bibitem{z:moy94}
V.~T. {Moy}, E.-L. {Florin}, and H.~E. {Gaub}, Science {\bf 266},  257  (1994).

\bibitem{z:lee94}
G.~U. {Lee}, D.~A. {Kidwell}, and R.~J. {Colton}, Langmuir {\bf 10},  354
  (1994).

\bibitem{nish95}
T. Nishizaka, H. Miyata, H. Yshikawa, S. Ishiwata
and K. Kinosita, Nature {\bf 377}, 251 (1995). 

\bibitem{bong99}
for a review, see: P. Bongrand, Rep. Prog. Phys. {\bf 62}, 921 (1999).


\bibitem{z:evan97d}
E. {Evans} and K. {Ritchie}, Biophys. J. {\bf 72},  1541  (1997).

\bibitem{z:izra97}
S. {Izrailev} {\it et~al.}, Biophys. J. {\bf 72},  1568  (1997).

\bibitem{shil98}
J. Shillcock and U. Seifert, Phys. Rev. E {\bf 57}, 7301 (1998).

\bibitem{merk99}
R. Merkel, P. Nassoy, A. Leung, K. Ritchie and E. Evans,
Nature {\bf 397}, 50 (1999).

\bibitem{sims99}
D. A. Simson, M. Strigl, M. Hohenadl, and R. Merkel,
Phys. Rev. Lett. {\bf 83}, 652 (1999).

\bibitem{z:grub96}
H. {Grubm{\"u}ller}, B. {Heymann}, and P. {Tavan}, Science {\bf 271},  997
  (1996).

\bibitem{bell78}
G.I. Bell, Science {\bf 200}, 618 (1978).

\bibitem{bell84}
G.I. Bell, M. Dembo, and P. Bongrand, Biophys. J. {\bf 45}, 1051 (1984).

\bibitem{zuck95}
D. Zuckerman and R. Bruinsma, Phys. Rev. Lett. {\bf 74}, 3900 (1995).

\bibitem{lipo96}
R. Lipowsky, Phys. Rev. Lett. {\bf 77}, 1652 (1996). 

\bibitem{nopp96}
D. A. Noppl-Simson and D. Needham, Biophys. J. {\bf 70}, 1391 (1996).

\bibitem{albe97}
A. Albersd\"orfer , T.  Feder, and E. Sackmann,
Biophys. J. {\bf 73}, 245 (1997). 

\bibitem{ches98}
S. E. Chesla, P. Selvaraj, and C. Zhu, Biophys. J. {\bf 75}, 1553 (1998).
                                         
\bibitem{demb88}
M. Dembo, D.C. Tornby, K. Saxmann, and D. Hammer,
Proc. R. Soc. Lond. B {\bf 234}, 55 (1988).

\bibitem{ra99} 
for an experimental study, see:
 H.J. Ra, C. Picart, H. Feng, H.L. Sweeney, and D.E. Discher,
J. Cell Sci. {\bf 112}, 1425 (1999).


\bibitem{hamm87}
D. A. Hammer and D. A. Lauffenburger,
Biophys. J. {\bf 52}, 475 (1987).

\bibitem{rief98}
M. Rief, J.M. Fernandez, and H.E. Gaub,
Phys. Rev. Lett. {\bf 81}, 4764 (1998).

\end{thebibliography}
\end{document}